\def\lae{\mathrel{<\kern-1.0em\lower0.9ex\hbox{$\sim$}}}
\def\gae{\mathrel{>\kern-1.0em\lower0.9ex\hbox{$\sim$}}}

\newcommand{\be}{\begin{equation}}
\newcommand{\ee}{\end{equation}}

\documentclass[twocolumn]{aastex6}
\usepackage{txfonts}
\usepackage{cases}

\shorttitle{Evidence for the Secondary Emission in TeV Blazars} \shortauthors{Zheng
et al.}

\begin{document}

\title{Evidence for the Secondary Emission as the Origin of Hard Spectra in TeV Blazars}
\author{Y. G. Zheng\altaffilmark{1} and T. Kang\altaffilmark{1}}
\altaffiltext{1}{Department of Physics, Yunnan Normal University, Kunming, 650092, China (E-mail:ynzyg@sohu.com)}

\begin{abstract}
We develop a model for a possible origin of hard very high energy spectra from a distant blazar. In the model, both the primary photons produced in the source and secondary photons produced outside the source contribute to the observed high energy $\gamma$-rays emission. That is, the primary photons are produced in the source through the synchrotron self-Compton (SSC) process, and the secondary photons are produced outside the source through high energy protons interaction with the background photons along the line of sight. We apply the model to a characteristic case was the very high energy (VHE) $\gamma$-ray emissions in distant blazar 1ES 1101-232. Assuming a suitable electron and proton spectra, we obtain excellent fits to observed spectra of distant blazar 1ES 1101-232. This indicated that the surprisingly low attenuation of high energy $\gamma$-rays, especially for the shape of the very high energy $\gamma$-rays tail of the observed spectra, can be explained by secondary $\gamma$-rays produced in interactions of cosmic-ray protons with background photons in the intergalactic space.
\end{abstract}

\keywords{BL Lacertae objects: individual
(1ES 1101-232)--radiation mechanisms:non-thermal }

\section{Introduction}
\label{sec:intro}
Blazars, a special class of active galactic nuclei (AGN), exhibit that the continuum emission, which arise from the jet emission taking place in AGN whose jet axis is closely aligned with the observer's line of the sight, is dominated by nonthermal emission as well as rapid and large amplitude variability(Urry \& Padovani 1995). The broad
spectral energy distributions (SED) from the radio to the $\gamma$-rays bands is dominated by two components, indicating two humps. It is widely admitted that the first hump is produced from electron synchrotron radiation, the peaks range from the infrared-optical up to X-rays regime for different blazars (Urry 1998). The second hump, the peaks in the GeV to TeV  $\gamma$-ray band, probably is produced from inverse Compton scattering of the
relativistic electrons either on the synchrotron photons(Maraschi et al. 1992) or on some other photon populations(Dermer et al. 1993; Sikora et al. 1994). As a open issue, high energy $\gamma$-ray is produced by mesons and leptons through the cascade initiated by proton-proton or proton-photon interactions(e.g. Mannheim \& Biermann 1992; Mannheim 1993; Phol \& Schlickeiser 2000; Aharonian 2000; M$\ddot{\rm u}$cke \& Protheroe 2001).

Observations of very high energy (VHE) $\gamma$-ray indicate that more than 40 blazars radiate $\gamma$-rays in the TeV energy regions(e.g. Aharonian et al. 2005; Cui 2007; Wagner 2008). It is believed that the primary TeV photons from the distant TeV blazars should exhibit clear signatures of absorption due to theirs interactions with the extragalactic background light (EBL) to produce electron-positron($e^{\pm}$)(e.g. Nikishov 1962; Gould \& Schreder 1966). However, the observed spectra do not show a sharp cutoff at energies around 1 TeV (Aharonian et al. 2006; Costamante et al. 2008; Acciari et a. 2009). One characteristic case was the very high energy (VHE) $\gamma$-ray emissions in distant blazar 1ES 1101-232, which detected by the high energy stereoscopic system (HESS) array of Cherenkov telescopes(Aharonian et al. 2006a; 2007a). The VHE $\gamma$-rays data result in very hard intrinsic spectra with a peak in the SED above 3 TeV, through corrected for absorption by the lowest level EBL (Aharonian et al. 2007a). A similar behavior has also been detected in other TeV blazars such as 1ES 0229+200 (Aharonian et al. 2007b), 1ES 0347-121 (Aharonian et al. 2007c) and Mkn 501 (Neronov et al. 2011).

Generally, the lack of absorption features either are simplest to assume that there is no absorption (Kifune 1999; Stecker \& Glashow 2001; De Angelis et al. 2009) or can be explained by the lower levels of EBL(Aharonian et al. 2006b; Mazin \& Raue 2007; Finke \& Razzaque 2009). Alternatively, the hard spectra can be expected, if the $\gamma$-rays from distant blazars may be dominated by secondary $\gamma$-rays produced along the line of sight by the interactions of cosmic rays protons with the background photons(Essey \& Kusenko 2010; Essey et al. 2010; Essey et al. 2011).

Active galactic nuclei (AGN) are believed to be the most powerful sources of both $\gamma$-rays and cosmic rays. The recently observed results by Cherenkov telescopes indicated that interactions of cosmic rays emitted by distant blazars with photon background along the line of sight can produce $\gamma$-rays (Essey \& Kusenko 2009).
Motivated by above arguments, in this paper, we study the possible origin of hard spectra in TeV blazars. The high energy emission from TeV blazars consists of two components: one, the primary $\gamma$-rays component, come from the source, and the other, secondary $\gamma$-rays component, come from proton interactions with the EBL photons along the line of sight.

Throughout the paper, we assume the Hubble constant $H_{0}=70$ km s$^{-1}$ Mpc$^{-1}$, the matter energy density $\Omega_{\rm M}=0.27$, the radiation energy density$\Omega_{\rm r}=0$, and the dimensionless cosmological constant $\Omega_{\Lambda}=0.73$.

\section{The Model}
\label{sec:model}
We basically follow the traditional synchrotron self-Compton (SSC) model to produce the primary $\gamma$-rays component and use the method given by Kelner \& Aharonial (2008) to produce the secondary $\gamma$-rays component.  now, we give a brief description on the model
\subsection{primary component produced in the source}
The homogeneous SSC radiation model is widely used for explaining the multi-wavelength energy spectra of blazars. The homogeneous SSC radiation model that we adopt assumes a spherical radiation region filled with extreme-relativistic electrons, in which the randomly-originated homogeneous magnetic field and constant electron number density exist. We adopt a broken power-law function with a sharp cut-off to describe the electron energy distribution in the radiation region:
\begin{equation}
N_{\rm e}(\gamma)=
\left\{
\begin{array}{ll}
K_{\rm 1}\gamma^{\rm -n_{\rm 1}}\;,\mbox{$\gamma_{\rm min}\le\gamma\le\gamma_{\rm b}$}\;;\\
K_{\rm 2}\gamma^{\rm -n_{\rm 2}}\;,\mbox{$\gamma_{\rm b}\le\gamma\le\gamma_{\rm cut}$}\;.
\end{array}
\right.
\label{eq:1}
\end{equation}
Where, $\gamma=E_{\rm e}/m_{\rm e}c^{\rm 2}$ is the Lorentz factor of electron and $K_{\rm 2}=K_{\rm 1}\gamma_{\rm b}^{\rm (n_{\rm 2}-n_{\rm 1})}$.

Based on the above electron number density $N(\gamma)$, we can use the formulae given by Katarzynski et al. (2001) to calculate the synchrotron intensity $I_{\rm s}(E_{\rm \gamma})$ and the intensity of self-Compton radiation $I_{\rm c}(E_{\rm \gamma})$(e.g. Zheng \& Zhang 2011), and then calculate the intrinsic photon spectrum at the observer's frame:
\begin{equation}
\frac{dN_{\rm \gamma}^{\rm int}}{dE_{\rm \gamma}}=\pi\frac{R^2}{d^2E_{\rm \gamma}^{2}}\delta^3(1+z)[I_{\rm s}(E_{\rm \gamma})+ I_{\rm c}(E_{\rm \gamma})]\;,
\end{equation}
where $d$ is the luminosity distance, $z$ is the redshift, and $\delta = [\Gamma(1 - \beta\cos \theta)]^{-1}$ is the Doppler factor where $\Gamma$ is the blob Lorentz factor, $\theta$ is the angle of the blob vector velocity to the line of sight and $\beta= v/c$. Since at high energies the Compton photons may produce pairs by interacting with the synchrotron photons, this process may be decrease the observed high energy radiation (Coppi \& Blandford 1990; Finke et al. 2009). Katarzynski et al. (2001) analyze the absorption effect due to pair-production inside the source, they found that its process is almost negligible. On the other hand, very high energy (VHE) $\gamma$-photons from the source are attenuated by photons from the extragalactic background light (EBL). Therefore, after taking the absorption effect, the flux density observed at the Earth becomes
\begin{equation}
\frac{dN_{\rm \gamma}^{\rm obs}}{dE_{\rm \gamma}}=\frac{dN_{\rm \gamma}^{\rm int}}{dE_{\rm \gamma}}\exp[-\tau(E_{\rm \gamma},z)]\;,
\end{equation}
where $\tau(E_{\rm \gamma},z)$ is the absorption optical depth due to interactions with the EBL (Kneiske et al. 2004; Dwek \& Krennrich 2005). In our calculation, we use the absorption optical depth which is deduced by the average EBL model in Dwek \& Krennrich (2005).

\subsection{secondary component produced outside the source}
AGNs are expected to accelerate cosmic rays to energies up to $\sim 10^{11}$ GeV. For energies below Greisen-Zatsepin-Kuzmin (GZK) cutoff of about 50 EeV (Greisen 1966; Zatsepin \& Kuzmin 1966), the cosmic rays can propagate cosmological distances without significant energy loss, then interact with EBL relatively close to Earth. The secondary $\gamma$-rays production by interactions of cosmic rays emitted by distant blazars with photon background along the line of sight mainly depend on the pion decay. Cosmic ray interactions with EBL should also produce neutrinos and electrons. We will concentrate on photons production and not discuss other particles in the proton-photon interaction processes.

Let $f_{\rm p}(E_{\rm p})$ and $f_{\rm ph}(\epsilon)$ be functions characterizing the energy distributions of initial protons and soft photons, the production rate of $\gamma$-rays can be obtained (Kelner \& Aharonian 2008):
\begin{equation}
\frac{dN_{\rm \gamma}}{dE_{\rm \gamma}}=\int f_{\rm p}(E_{\rm p})f_{\rm ph}(\epsilon)\Phi_{\rm \gamma}(\eta, x)\frac{dE_{\rm p}}{E_{\rm p}}d\epsilon\;.
\end{equation}
Where, $\eta=4\epsilon E_{\rm p}/(m_{\rm p}^{2}c^{4})$, $x=E_{\rm \gamma}/E_{\rm p}$, and $\Phi_{\rm \gamma}(\eta, x)$ is a piecewise function of two variables. According to the results on photo-meson processes, what are obtained using numerical simulations of proton-photon interactions based on the public available Monte-Carlo code SOPHIA (M$\ddot{\rm u}$cke et al. 2000), Kelner \& Aharonian (2008) give an approximate analytical presentations. Namely,
\begin{equation}
\Phi_{\rm \gamma}(\eta,x)=\left\{
\begin{array}{lc}
B_{\rm \gamma}^{}\,\exp\!\left\{-s_{\rm \gamma}^{}
\left[\ln\!\left(\frac{x}{x_-}\right)\right]^{\delta_{\rm \gamma}}\right\}\nonumber\\ \times
\left[\ln\!\left(\frac{2}{1+y^2}\right)\right]^{2.5+0.4\ln(\eta/\eta_{0})}\,,& x_{-}<x<x_{+}\;,\\[3pt]
B_{\rm \gamma}\left(\ln 2\right)^{2.5+0.4\ln(\eta/\eta_{0})}\,,& x\leqslant x_{-}\;,\\[3pt]
0\,,& x\geqslant x_{+}\;.\\[3pt]
\end{array}
\right.
\end{equation}
Where, $y=(x-x_{-})/(x_{+}-x_{-})\,$, $\eta_{0}$ relate to the proton mass $m_{\rm p}$ and the $\pi$-meson mass $m_{\rm \pi}$, $\eta_{0}=2m_{\rm \pi}/m_{\rm p}+m_{\rm \pi}^{2}/m_{\rm p}^{2}\,$, and $x_{\pm}=\frac{1}{2(1+\eta)}\,\left[\eta+r^{2}\pm\sqrt{(\eta-r^{2}-2r)\,(\eta-r^{2}+2r)}\right]\,$
with $r=m_{\rm \pi}/m_{\rm p}\approx 0.146\,$. All three parameters $B_{\rm \gamma}$, $s_{\rm \gamma}$, and $\delta_{\rm \gamma}$ used in above presentation are functions of $\eta$. The numerical values of these parameters are be given in Kelner \& Aharonian (2008).

Instead of integrating Eq (4) over $d\epsilon$, it is more convenient to perform integration of the equation over $d\eta$. This allow us to rewrite spectra of $\gamma$-ray photons in the form
\begin{equation}
\frac{dN_{\gamma}}{dE_{\gamma}}=\int_{\eta_{0}}^{\eta_{\rm max}}H(\eta,E)\,d\eta\,.
\end{equation}
where
\begin{equation}
H(\eta,E_{\gamma})=\frac{m_{p}^{2}c^{4}}{4}\!\int_{E}^{\infty}\!\frac{dE_{p}}{E_{p}^{2}}\,
f_{p}(E_{p})\, f_{\rm ph}\!\left(\frac{\eta m_{p}^{2}c^{4}}{4E_{p}}\right)\!
\Phi_{\gamma}\!\left(\eta,\frac{E_{\gamma}}{E_{p}}\right)\!.
\end{equation}

\section{Apply to the Hard spctra in 1ES 1101-232}
\label{sec:apply}
1ES 1101-232 resides in an elliptical host galaxy at a redshift of $z=0.186$(Remillard et al. 1989; Falomo et al. 1994). The source has been classified as a high frequency peaked BL BLac object (HBL), because of he dominance of synchrotron emission in the X-ray band(Donato et al. 2001). In previously published SSC frame, the broadband characteristics of 1ES 1101-232 indicated that the inverse Compton (IC) peak was generally expected to be around 100 GeV (e.g. Wolter et al. 2000; Costamante \& Ghisellini 2002). But, the new observational results with the HESS in 2004, especially in 2005 indicated that the source exhibits a hard intrinsic spectra with a peak in the SED above 3 TeV, through corrected for absorption by the lowest level EBL (Aharonian et al. 2007a). Using the model in $\S$2, we can calculate the TeV $\gamma$-rays spectra in the source (primary component) and outside the source (secondary component), respectively. Then a hard intrinsic spectra of the source can be produced. The two epochs will be considered independently.

In order to do so, first we search for the primary $\gamma$-ray component in the one-zone SSC frame. Assuming the electron Lorentz factors $\gamma_{\rm min}$, $\gamma_{\rm break}$, $\gamma_{\rm max}$ are in two epoches identical. We calculate the high energy electron distribution with a broken power law between $\gamma_{\rm min}=1$ and $\gamma_{\rm max}=8.0\times10^{8}$ with a break at $\gamma_{\rm break}=9.0\times10^{4}$, where, for 2004 observed data, the density normalization $\rm K=1200~cm^{-3}$, the energy index of the particles between $\gamma_{\rm min}$ and $\gamma_{\rm break}$ is set to $\rm n_{1}=2$ and the energy index of the particles between $\gamma_{\rm break}$ and $\gamma_{\rm max}$ is set to $\rm n_{2}=3.5$, and for 2005 observed data, the density normalization $\rm K=3.5~cm^{-3}$, the energy index $\rm n_{1}$ set to 1.8, and $\rm n_{2}$ set to 4.05. The parameters are used as follows. In the first observation epoch, the magnetic field strength is $\rm B=0.55$ G, the emission
region size is $\rm R=1.28\times10^{16}$ cm, and the Doppler factor is $\rm \delta=10.5$. In the later observation epoch, in order to obtain good fits, the magnetic field strength is $\rm B=0.15$ G, the emission region size is $\rm R=1.65\times10^{17}$ cm, and the Doppler factor is $\rm \delta=10$. All the physical parameters of the one-zone SSC spectra are listed in table 1.

 \begin{deluxetable}{lll}
 \tablecaption{Physical parameters of the one-zone SSC model spectra}
 \tablewidth{0pt}
 \tablehead{
 \colhead{parameters} & \colhead{2004} & \colhead{2005}
}
 \startdata
 $\gamma_{\rm min}$  & 1.0 & 1.0 \\
 $\gamma_{\rm break}$  & $9.0\times10^{4}$  & $9.0\times10^{4}$ \\
 $\gamma_{\rm max}$  & $8.0\times10^{8}$  & $8.0\times10^{8}$  \\
 K~[$\rm cm^{-3}$] & 1200 & 3.5\\
 $\rm n_{1}$ &2.0 & 1.8 \\
 $\rm n_{2}$ & 3.5 & 4.05 \\
 B~[G]   &  0.55 & 0.15  \\
 R~[cm] & $1.28\times10^{16}$ & $1.65\times10^{17}$ \\
 $\delta$ &10.5 & 10 \\
\enddata
 \end{deluxetable}

We assume that relativistic electrons are in the steady state during the observational epoch. Therefore, we can
calculate the X-ray/TeV $\gamma$-ray spectrum in the one-zone SSC model using the broken power law electron spectrum. In Fig.\ref{Fig:1}, we show predicted spectrum from X-ray to TeV $\gamma$-ray bands (blue solid curve). For comparison, observed data of 1ES 1101-232 at X-ray band and TeV band on the June 5-10, 2004 and March 5-16, 2005
(Aharonian et al. 2007a) are also shown, respectively. It can be seen that the lower energy observed data can be reproduced in the SSC model.

\begin{figure}
\vspace{0.0cm}
\label{Fig:1}
\epsscale{1.0} \plotone{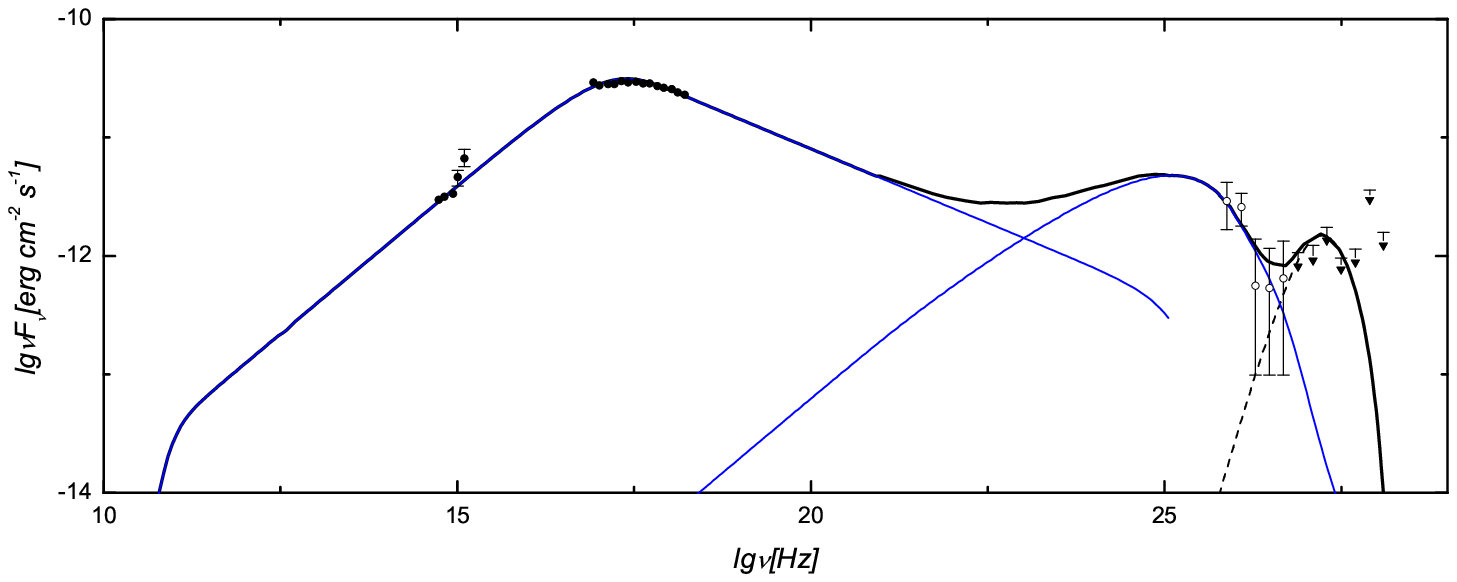}
\vspace{0.0cm}
\plotone{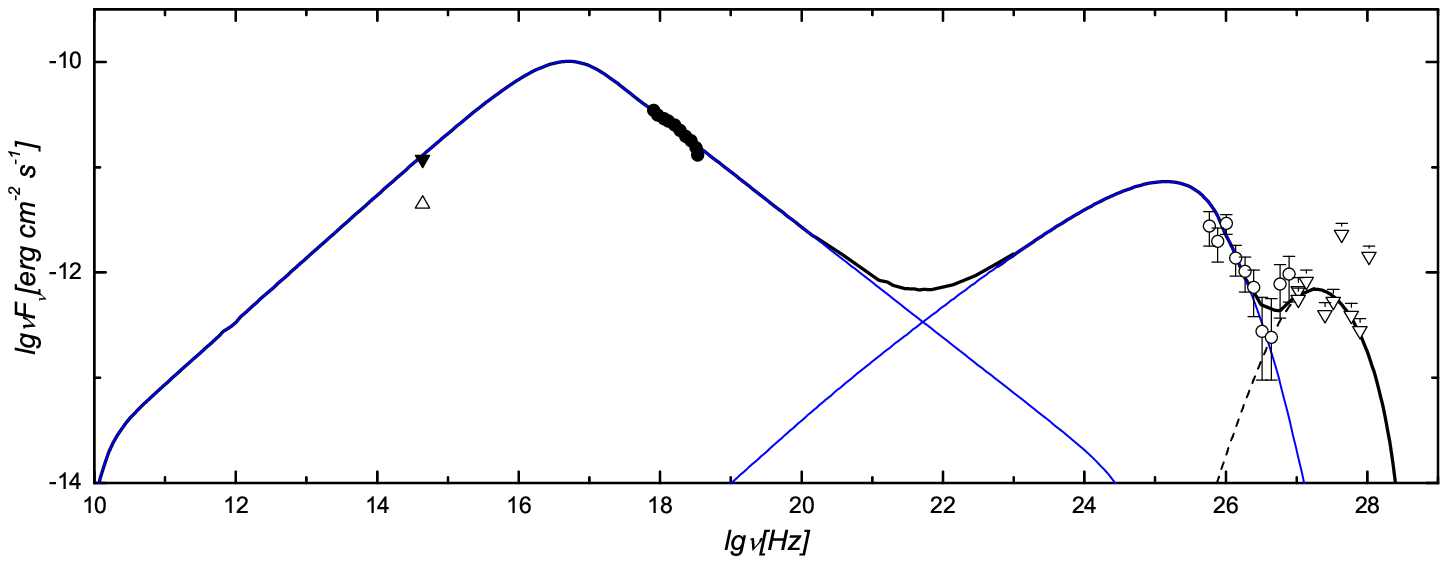}
\caption{Comparisons of predicted multi-wavelength spectra with observed data of 1ES 1101-232 on the June 5-10, 2004 (top panel) and on the March 5-16, 2005 (bottom panel). Blue solid curves represent the primary component (SSC) spectra, black dash curve represent the secondary component (or proton-photon interaction) spectra, and black solid curve represent the total spectra, respectively. Observed data come from Aharonian et al. (2007a).}
  \label{Fig:1}
\end{figure}

We now consider the hard spectra properties of 1ES 1101-232 in 2004 June and 2005 March, especially for the shape of the very high energy $\gamma$-ray tail of the observed spectra. In order to do it, we calculate the spectra of secondary $\gamma$-rays numerically using the EBL spectra $n(z, \epsilon)$ which is deduced by the average EBL model in Dwek \& Krennrich (2005) at the redshift $z$. The energy losses are due to production of pions in proton-photon ($p-\gamma$) interactions with the EBL photons. This process depends on the proton injection spectrum, which we parameterize by a constant power-law exponent $\alpha$ and maximal energy $\rm E_{max}$:
\begin{equation}
f_{\rm p}(E_{\rm p})=N_{0}E_{\rm p}^{\alpha}exp\left(-\frac{E_{\rm p}}{E_{\rm max}}\right)\;.
\end{equation}
Where the normalization coefficient $N_{0}$ is determined from the condition
\begin{equation}
\int E_{\rm p}f_{\rm p}(E_{\rm p})dE_{\rm p}=1~erg~cm^{-3}\;.
\end{equation}

Assuming the minimum energy of the injection proton to be of the order of 0.145 GeV, which is a kinematic threshold of the photomeson production processes in proton-photon interactions in two epoches identical. We calculate the high energy injection proton spectra with the $N_{0}=10.0~\rm erg~cm^{-3}$, $E_{\rm p~max}=1.28\times10^{-3}$ GeV for the 2004 observed data, and with the $N_{0}=2.0~\rm erg~cm^{-3}$, $E_{\rm p~max}=3.02\times10^{-3}$ GeV for the 2005 observed data, respectively. Under the above proton spectra, we reproduce the observed TeV photon spectrum (dashed curve) of 1ES 1101-232 on the on the June 5-10, 2004 and March 5-16, 2005 in Fig.\ref{Fig:1}, respectively. It can be seen that the observed hard spectra properties of 1ES 1101-232 in 2004 June and 2005 March, especially for the shape of the very high energy $\gamma$-ray tail of the observed spectra, can be reproduced in our model.

 \begin{deluxetable}{lcc}
 \tablecaption{Physical parameters of the proton injection spectrum}
 \tablewidth{0pt}
 \tablehead{
 \colhead{parameters} & \colhead{2004} & \colhead{2005}
}
 \startdata
 $E_{\rm p~min}$~[GeV] & 0.488 & 0.488 \\
 $E_{\rm p~max}$~[GeV]  & $1.28\times10^{5}$  & $3.02\times10^{5}$  \\
 $N_{0}$~[$\rm erg~cm^{-3}$] & 10.0 & 2.0\\
 $\alpha$ & -2.0 & -2.0\\
\enddata
 \end{deluxetable}

\section{Discussion and conclusions}
\label{sec:discussion}
Generally, the proton-proton ($p-p$) interactions do not offer efficient $\gamma$-rays production mechanisms in the jet. This mechanism could be effectively realized only in a scenario assuming that $\gamma$-rays is produced in dense gas clouds that move across the jet (e.g. Morrison et al. 1984; Dar \& Laor 1997), such as, in order to interpret the reported TeV flares of Markarian 501 by $\rm \pi^{0}$-decay $\gamma$-rays produced at $p-p$ interactions, for any reasonable acceleration power of protons $\rm L_{p}\le 10^{45}~erg~s^{-1}$, the density of the thermal plasma in the jet should exceed $\rm 10^{6}~cm^{-3}$ (Aharonian 2000). On the other hand, at the conditions of existence of extremely high energy, $\rm E>10^{19}~eV$, and in presence of large magnetic field, $\rm B>>1~G$, the synchrotron radiation of the protons becomes a very effective channel of production of high energy $\gamma$-rays. In our calculations, in order to reproduce the observed spectra of distant blazar 1ES 1101-232, we adopt a lower protons energy and magnetic field. these postulations lead to a longer ages of proton with $\rm t_{sy}=4.5\times10^{4}B_{100}^{-2}E_{19}^{-1}~s$ (Aharonian 2000), where $\rm B_{100}=B/100~G$, $\rm E_{19}=E/10^{19}~eV$, and fainter radiation in the jet.

The proton induced cascade process (Mannheim 1993; 1996) is another attractive possibility for production of high energy $\gamma$-rays. This process relates the observed $\gamma$-rays radiation to the development of pair cascades in the jet triggered by secondary photo-meson products produced at interactions of accelerated protons with low frequency synchrotron radiation in the source or EBL photons outside the source. For a low energy target photon field, the photo-meson cooling time of protons can be estimated using the approximate formula $\rm t_{p\gamma}\sim1/<\sigma_{p\gamma}K_{p\gamma}>cn_{ph}(\nu>\nu_{th})$ (Begelman et al. 1990), where $\rm <\sigma_{p\gamma}K_{p\gamma}>\sim 0.7\times10^{-28}~cm^{2}$ is the photo-meson production cross section and inelasticity parameter averaged over the resonant energy range (e.g. Stecker 1968; M$\ddot{\rm u}$cke et al. 1999). We simply approximate the broad synchrotron spectral component by a power-law function with the energy-flux index $\rm \alpha=1$ and denoting its luminosity by $\rm L_{s}$, we have $\rm n_{ph}(\nu>\nu_{th})\sim L_{s}E_{p}/(4\pi m_{\pi}m_{e}c^{5}R^{3}\delta^{4})$ (e.g. Sikora 2010). Thus, for the parameters of 1ES 1101-232, the photo-meson cooling time can not be significantly less than light travel timescales $\rm R/c\sim 10^{7}~s$. We argue that the uncooled protons can escape from the emission region, and then interact with the background photons along the line of sight.

In this paper, we develop a model for a possible origin of hard very high energy spectra from a distant blazar. Though, several models which could explain very hard intrinsic blazar spectra in the $\gamma$-ray band have already been proposed (Katarzynski et al. 2006; B$\rm \ddot{o}$ttcher et al. 2008; Aharonian et al. 2008; Lefa et al. 2011; Yan et al. 2012). In the model, both the primary photons produced in the source and secondary photons produced outside the source contribute to the observed high energy $\gamma$-rays emission. That is, the primary photons are produced in the source through the SSC process, and the secondary photons are produced outside the source through high energy protons interaction with the background photons along the line of sight. Assuming a suitable electron and proton spectra, we obtain excellent fits to observed spectra of distant blazar 1ES 1101-232. This indicated that the surprisingly low attenuation of high energy $\gamma$-rays, especially for the shape of the very high energy $\gamma$-rays tail of the observed spectra, can be explained by secondary $\gamma$-rays produced in interactions of cosmic-ray protons with background photons in the intergalactic space(Essey \& Kusenko 2010; Essey et al. 2010; 2011).

The properties of the model mentioned above should have been testable in the multi-wavelength observations on TeV blazar. Costamante (2012) argue that in several cases we have already seen the superposition of two different emission components at high electron energies, with a new components emerging over a previous SED. Since, in our case, the secondary $\gamma$-rays are produced outside of the host galaxy, this expects to harder spectra in TeV energy band than GeV energy band. These should be verified in the future multi-wavelength observations. Otherwise, the neutrino populations can be expected in $\rm p\gamma$ interactions, we leave this possibility for next work and IceCube observations.
\section*{Acknowledgments}
This work is partially supported by the National Natural Science
Foundation of China under grants 11178019 and the
Natural Science Foundation of Yunnan Province under grants
2011FB041. This work is also supported by the Science
Foundation of Yunnan educational department (grant 2012Z016).


\end{document}